\documentclass[preprint, letterpaper, 12pt]{aastex}
\usepackage[letterpaper]{geometry}
\usepackage{helvet}
\usepackage{natbib}
\usepackage[colorlinks,urlcolor=blue,citecolor=blue,linkcolor=blue]{hyperref} 
\usepackage{xspace}

\newcommand{\kms}{~km~s$^{-1}$\xspace}
\newcommand{\nthp}{N$_2$H$^{+}$\xspace}

\begin{document}
\title{CARMA Large Area Star Formation Survey: Observational Analysis of Filaments in the Serpens South Molecular Cloud}
\author{M. Fern\'andez-L\'opez\altaffilmark{1,2}, H.~G. Arce\altaffilmark{3}, L. Looney\altaffilmark{1}, L.~G. Mundy\altaffilmark{4}, S. Storm\altaffilmark{4}, P.~J. Teuben\altaffilmark{4}, K.~Lee\altaffilmark{4,1}, D.~Segura-Cox\altaffilmark{1}, A.~Isella\altaffilmark{5}, J.~J.~Tobin\altaffilmark{6}, E.~Rosolowsky\altaffilmark{7,8}, A.~Plunkett\altaffilmark{3}, W.~Kwon\altaffilmark{9}, J.~Kauffmann\altaffilmark{5}, E.~Ostriker\altaffilmark{10}, K.~Tassis\altaffilmark{11,12}, Y.~L. Shirley\altaffilmark{13,14}, M. Pound\altaffilmark{4}}

\altaffiltext{1}{Department of Astronomy, University of Illinois at Urbana--Champaign, 1002 West Green Street, Urbana, IL 61801, USA; manferna@gmail.com}   
\altaffiltext{2}{Current address: Instituto Argentino de Radioastronom{\'{\i}}a, CCT-La Plata (CONICET), C.C.5, 1894, Villa Elisa, Argentina}
\altaffiltext{3}{Department of Astronomy,Yale University, P.O.~Box 208101, New Haven, CT 06520-8101, USA}   
\altaffiltext{4}{Department of Astronomy, University of Maryland, College Park, MD 20742, USA}   
\altaffiltext{5}{Astronomy Department, California Institute of Technology, 1200 East California Blvd., Pasadena, CA 91125, USA}   
\altaffiltext{6}{National Radio Astronomy Observatory, Charlottesville, VA 22903, USA}
\altaffiltext{7}{University of British Columbia, Okanagan Campus, Departments of Physics and Statistics, 3333 University Way, Kelowna BC V1V 1V7, Canada}
\altaffiltext{8}{University of Alberta, Department of Physics, 4-181 CCIS, Edmonton AB T6G 2E1, Canada}
\altaffiltext{9}{SRON Netherlands Institute for Space Research, Landleven 12, 9747 AD Groningen, The Netherlands}
\altaffiltext{10}{Department of Astrophysical Sciences, Princeton University, Princeton, NJ 08544, USA} 
\altaffiltext{11}{Department of Physics and Institute of Theoretical \& Computational Physics, University of Crete, PO Box 2208, GR-710 03, Heraklion, Crete, Greece}
\altaffiltext{12}{Foundation for Research and Technology - Hellas, IESL, Voutes, 7110 Heraklion, Greece}
\altaffiltext{13}{Steward Observatory, 933 North Cherry Avenue, Tucson, AZ 85721, USA}
\altaffiltext{14}{Adjunct Astronomer, The National Radio Astronomy Observatory}

\begin{abstract}
We present the \nthp($J=1\rightarrow0$) map of the Serpens South molecular cloud obtained as part of the CARMA Large Area Star Formation Survey (CLASSy). The observations cover 250 square arcminutes and fully sample structures from 3000~AU to 3~pc with a velocity resolution of 0.16\kms, and they can be used to constrain the origin and evolution of molecular cloud filaments.
The spatial distribution of the \nthp emission is characterized by long filaments that resemble those observed in the dust continuum emission by {\it Herschel}. However, the gas filaments are typically narrower such that, in some cases, two or three quasi-parallel \nthp filaments comprise a single observed dust continuum filament. The difference between the dust and gas filament widths casts doubt on \textit{Herschel} ability to resolve the Serpens South filaments. Some molecular filaments show velocity gradients along their major axis, and two are characterized by a steep velocity gradient in the direction perpendicular to the filament axis.  
The observed velocity gradient along one of these filaments was previously postulated as evidence for mass infall toward the central cluster, but these kind of gradients can be interpreted as projection of large-scale turbulence. 
\end{abstract}

\section{INTRODUCTION}
Recent observations have shown that dense molecular filaments can host the collapse of protostellar cores (e.g., Myers~2009; Andr{\'e} et al.~2010).  Moreover, clusters of massive pre-stellar cores have been found at the junctions of filaments (Schneider et al.~2012; Ysard et al.~2013). The ubiquitous filaments and filamentary hubs in molecular clouds are therefore thought to play a fundamental role in converting the quiescent gas into protostellar cores.  
However, the origin and ultimate fate of the filaments and their role in the star formation process are still not well known. In this letter we present high angular resolution \nthp data, taken as part of the CARMA Large Area Star Formation Survey (CLASSy), that highlight the filamentary structure of the Serpens South cloud.

{\it Herschel} observations of cold dust emission have renewed interest in molecular cloud filaments, thanks in part to the large areas covered. These large-scale images have allowed for the analysis of parsec-scale structures at resolutions between 18$\arcsec$ and 37$\arcsec$ at various submillimeter wavelengths. More recently, {\it Herschel} data have been compared with gas kinematics provided by observations of molecular line emission. These studies have been typically conducted using single-dish observations, as they provide a good compromise between the angular resolution and the time required to map large regions (cf., Busquet et al.~2013). These observations have revealed {\it bundles} of filaments with slightly different radial velocities inside the larger filaments identified in the {\it Herschel} images (Hacar \& Tafalla 2011; Hacar et al.~2013; Jimenez-Serra et al.~2014). 
Furthermore, they have revealed velocity gradients along filaments, which have been interpreted as material flowing through the filaments into their junctions, where filament hubs host active star-forming regions (Kirk et al.~2013; Peretto et al.~2014).

CLASSy is a CARMA Key Project that has mapped five fields covering hundreds of square arcminutes in the nearby clouds of Perseus and Serpens (Storm et al.~2014). The molecular line maps are constructed using interferometric and autocorrelation data in order to create fully spatially sampled images, tracing structures from parsecs down to 3000-AU scales (synthesized beam $\sim7\arcsec$). One CLASSy region, the Serpens South cloud\footnote{We assume Serpens South is part of the Serpens Main complex and adopt a distance of 415~pc (Dzib et al. 2010)}, has received considerable attention since its recent discovery with {\it Spitzer} (Gutermuth et al.~2008) because of its remarkable filamentary structure (see Figure \ref{serps}), as well as its extreme youth and high protostellar density. Multi-wavelength observations indicate there are about 160 dust continuum sources at different evolutionary stages (Maury et al.~2011). The high count of protostars compared to Class II YSOs indicates this cluster is very young (a few $10^5$~yr) and active. Because of its youth, the star formation activity has not had much time to impact the natal cloud structure. The Serpens South cluster region is thus an ideal place to study the dense gas kinematics and morphology of filaments before significant gas dispersal has occurred. In the last year, a few contributions addressing the large-scale kinematics and molecular content of this region have been published (Friesen et al.~2013; Kirk et al.~2013; Tanaka et al.~2013). However, the angular resolution of all these observations is coarser than $20\arcsec$ while the new CLASSy data is $7\arcsec$.

\section{CARMA OBSERVATIONS}
The Serpens South molecular cloud was observed in a campaign between August 2012 and August 2013 with the CARMA 23-element mode in the two most compact configurations, E and D, covering an area of 250~square arcminutes. The autocorrelation data from the 10~m dishes were also used to make single-dish images in order to recover the large-scale line emission filtered out by the interferometer. The final mosaic is composed of 980 Nyquist sampled pointing observed in 42 tracks with good weather conditions totaling 221 hours of observations. The correlator in CARMA 23-element mode has four bands. Three bands were set to three different spectral lines and one band observed the continuum at 92.8~GHz. In this letter we only discuss the \nthp($J=1\rightarrow0$) data ($\nu_0=93.173704$~GHz), taken in a 8~MHz band of 159 channels, each 0.16~\kms wide (see Fern\'andez-L\'opez et al. in prep. for the rest of the data). Details on observations and processing of CLASSy data are provided in Storm et al.~(2014).

\section{RESULTS AND DISCUSSION}
\subsection{Morphology of the \nthp Filaments}
\begin{figure}
\includegraphics[scale=0.83]{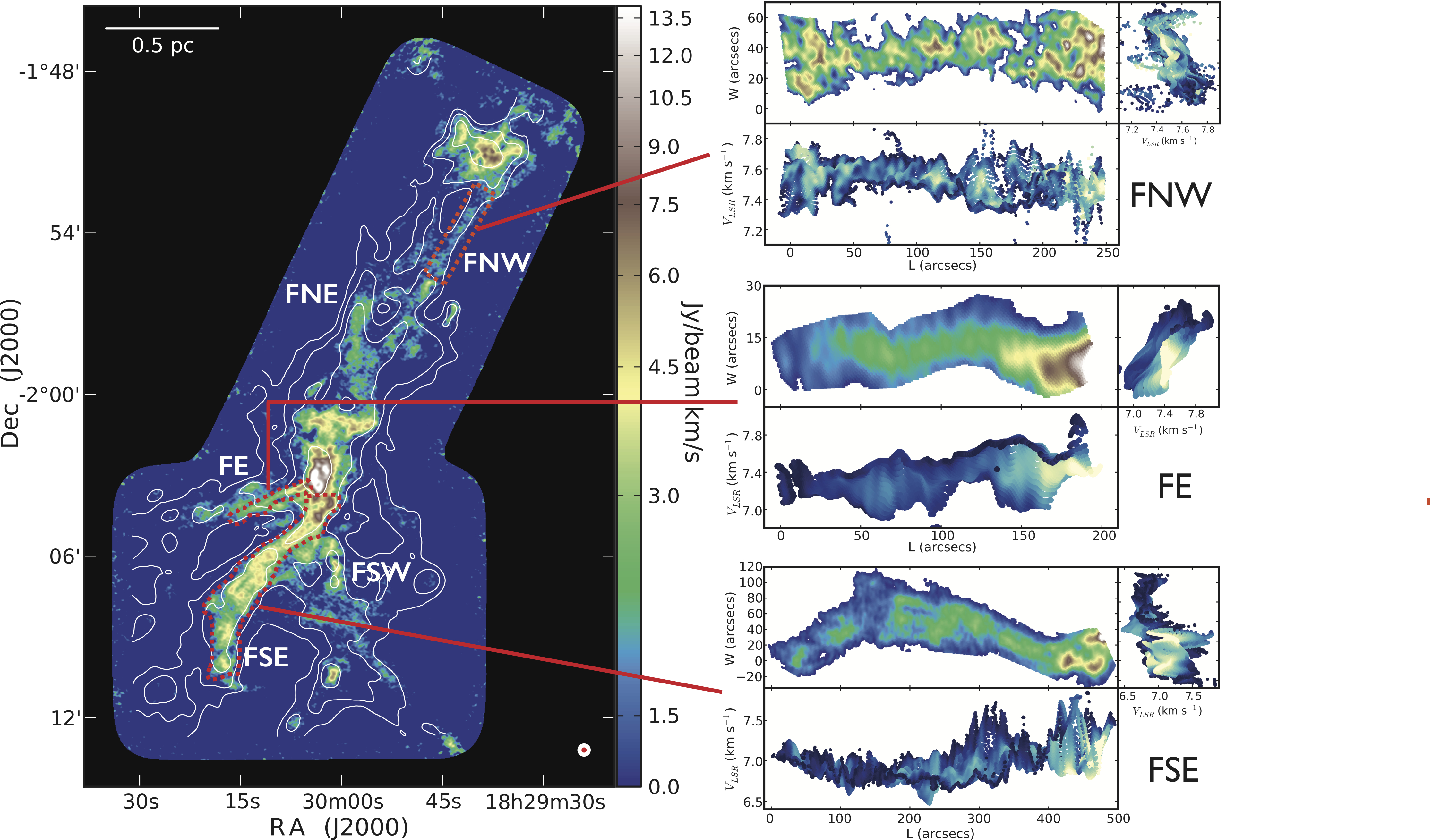}
\caption{Serpens South CARMA \nthp($J=1\rightarrow0$) integrated intensity image (color scale) overlaid with 350~$\mu$m {\it Herschel} data (white contours). The three external panels show the integrated intensity and position-velocity diagrams for the FNW, FE and FSE filaments. In these panels the filaments have been rotated to make them look horizontal. The position-velocity diagrams (along and across each filament) show a color scheme with brighter colors corresponding to higher intensities. The FE and FSE filaments show smooth velocity gradients along the filament length, while the FNW and FE filaments show velocity gradients across the filament. 
}
\label{serps}
\end{figure}

The large-scale structure of the Serpens South molecular cloud is defined by parsec-scale filamentary structures around a central hub where a dense cluster of protostars lies. Figure \ref{serps} shows a color map of the CLASSy \nthp($J=1\rightarrow0$) integrated emission, where the white contours represent the {\it Herschel} 350~$\mu$m continuum emission (25$\arcsec$ beam). The figure shows that the \nthp emission generally follows the dust emission detected with {\it Herschel} in the submillimeter, tracing the cold and dense gas in elongated structures commonly seen as filaments, and identified in this case by visually inspecting the \nthp integrated intensity map.
The most prominent filament runs southeast of the hub (which we label FSE); there are also filaments  due east (FE) and southwest (FSW) (Figure \ref{serps}). 
North of the hub, the high-resolution CLASSy \nthp observations clearly separate the main dusty filamentary structure into at least two separate filaments not resolved in the {\it Herschel} images: filaments FNE and FNW (Figure \ref{serps}).  
Although most of the filaments are seen in both \nthp and dust emission, there are a few filaments only detected in dust (e.g.,  southwest or northeast structures on the map). We speculate that this can be due to the lower density of these filaments and/or a decrease in the \nthp abundance due to the reaction with increased CO abundances in regions of lower density and/or higher temperature (e.g., Bergin et al.~2001, 2002).

\begin{figure}
\centering
\begin{minipage}{.45\textwidth}
\centering
\includegraphics[scale=0.45]{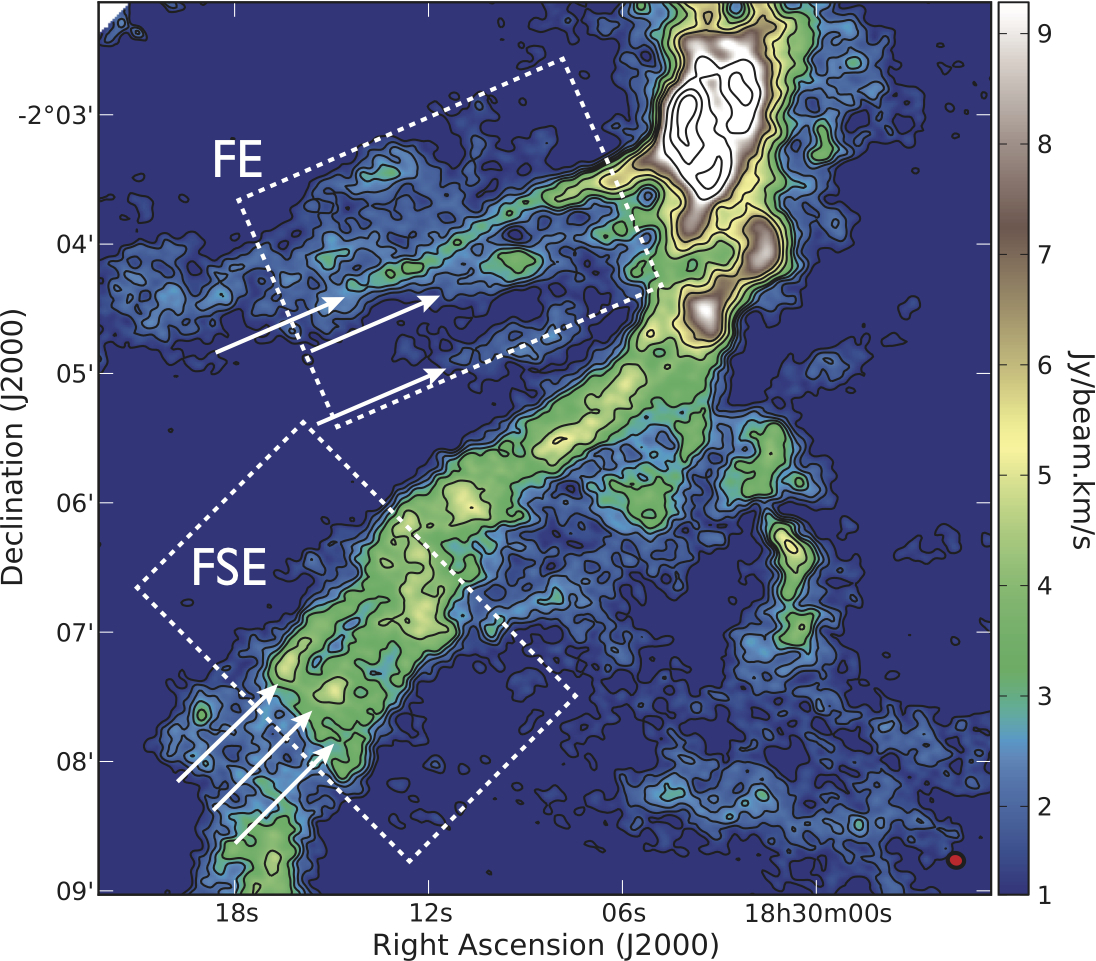}
\end{minipage}
\hspace{1.5cm}
\begin{minipage}{.45\textwidth}
\centering
\includegraphics[scale=0.45]{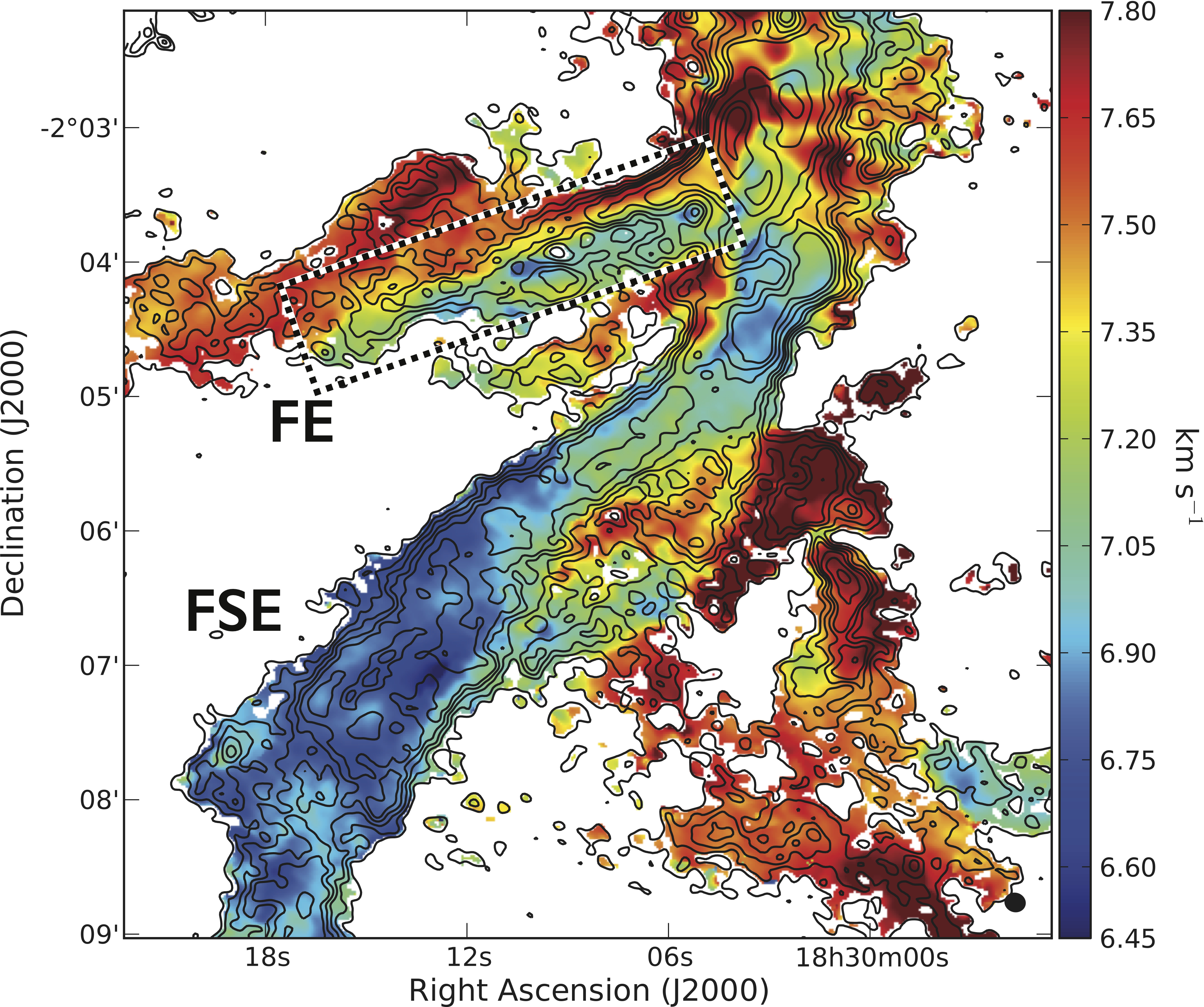}
\end{minipage}
\caption{{\bf a)} CARMA \nthp($J=1\rightarrow0$) integrated intensity image of a portion of Serpens South that includes the FSE and FE filaments. Contours are at 1.2,~1.7,~2.2,~2.7,~3.2,~4.2,~5.2,~6.2,~10.2,~14.2 and 18.2~Jy~beam$^{-1}$~\kms; the rms noise level is 0.3~Jy~beam$^{-1}$~\kms. The synthesized beam ($7\farcs7\times6\farcs8$) is shown in the bottom corner (red ellipse). The two boxes mark the location from which the intensity profile of Figure \ref{intcuts} were taken. The arrows show several quasi-parallel filament paths. {\bf b)} \nthp($J=1\rightarrow0$) velocity centroid (color scale) with contours representing the integrated intensity. The velocity gradient perpendicular to the FE filament is clearly seen. The box marks the velocity gradient across filament FE.}
\label{moms}
\end{figure}

\begin{figure}
\begin{minipage}{.33\textwidth}
\includegraphics[scale=0.33]{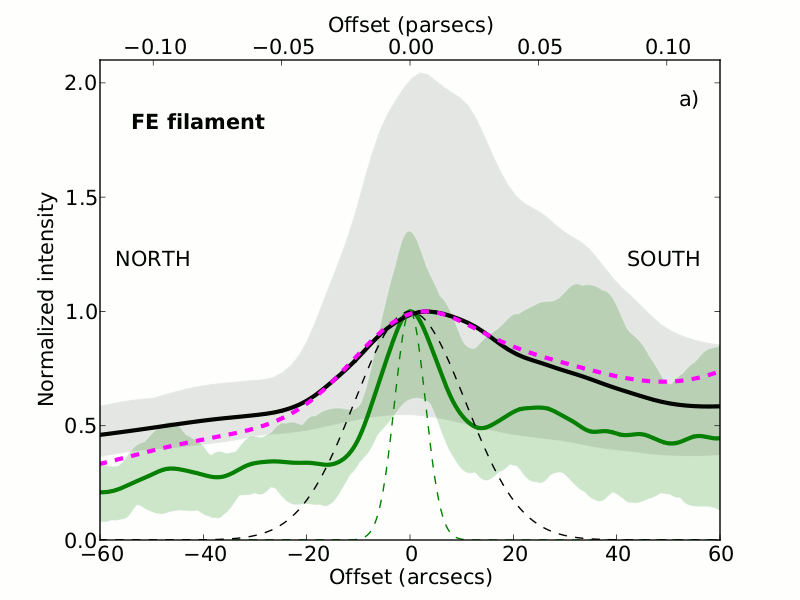}
\end{minipage}
\hspace{3.0cm}
\begin{minipage}{.33\textwidth}
\includegraphics[scale=0.33]{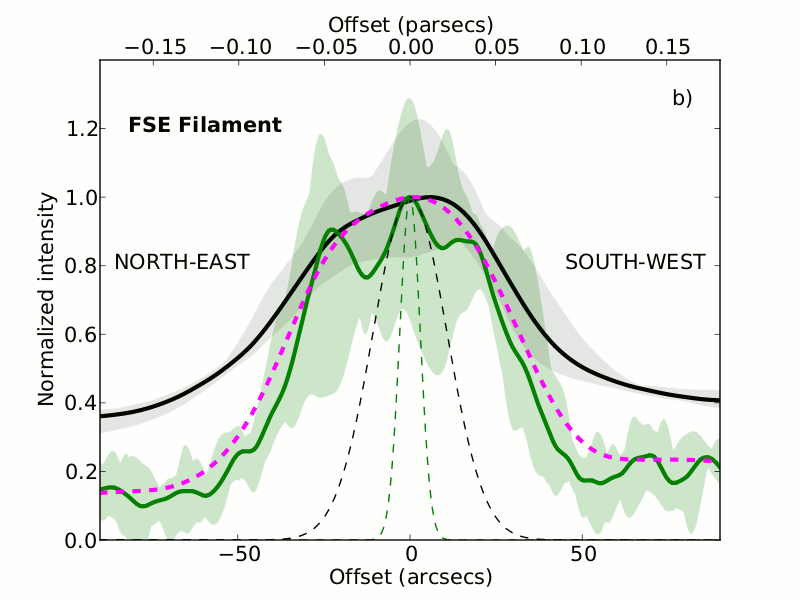}
\end{minipage}
\caption{{\bf a)} Normalized intensity cuts across the the FE filament in the CARMA \nthp($J=1\rightarrow0$) and 350~$\mu$m images (green and black curves and shades, respectively). The solid lines represent the average to each cut; the shades represent the range of values at each offset position centered at the crest of the filament.  A Gaussian fit to each curve shows that the \nthp width is 1/3 of the {\it Herschel} dust width. Green and black dashed lines show the CARMA and {\it Herschel} Gaussian beams. Violet dashed lines show the \nthp average convolved with the Herschel beam. {\bf b)} Same for the FSE filament. The green profile shows a substructure of two or three peaks (green line) inside the dust filament (black line). Both profiles were made using the emission inside the boxes shown in Figure \ref{moms}a.}
\label{intcuts}
\end{figure}

Analysis of the \nthp CLASSy data provides a detailed view of the morphology and kinematics of these filaments. In this letter we focus on three filaments (FE, FSE and FNW, Figure \ref{serps}). The \nthp filaments are seen as elongated structures, with slightly curved morphologies (Figure \ref{moms}a). The \nthp filament length is similar to that of its associated dusty filament, but its path is often separated by lower intensity emission. The widths of the \nthp filaments are narrower than those estimated from the {\it Herschel} data. In Figure \ref{intcuts}a we compare the width of the dust and the \nthp($J=1\rightarrow0$) integrated intensity emission of the FE filament by constructing an average intensity profile perpendicular to the filament main axis. A Gaussian fit to the \nthp intensity profile of this filament gives a FWHM$=0.035$~pc, while a fit to the 350~$\mu$m dust continuum emission gives a FWHM$=0.110$~pc. A similar analysis for the FN filament (not shown) also shows that the gas filament width is about one-half of the dusty filament width. 

{\it Herschel} observations of molecular cloud filaments find that the filament width distribution,  for a sample of 278 filaments in different molecular clouds with varying distances, has a strong peak at $0.09\pm0.04$~pc (Arzoumanian et al.~2011; Andr\'e et al.~2013). Andr\'e et al. conclude that this characteristic width is a \lq\lq quasi-universal\rq\rq~value, which implies that $0.1$~pc must be an intrinsic scale of the filament formation process.  
Figure \ref{intcuts}a shows how the \nthp filament profile convolved (violet dashed line) with the 350~$\mu$m {\it Herschel} beam matches quite well the dust profile (black line), while the $7\arcsec$ CLASSy profile (green curve) shows a prominent centrally peaked filament. 
Why is the \nthp emission more peaked than the dust continuum?

There are key differences in how the dust continuum and \nthp emission trace the material in filaments. The dust continuum emission at 300-500~$\mu$m increases with increasing temperature and column density, with no dependence on density. The \nthp emission depends on density, \nthp column density, and gas temperature through the line excitation and opacity. Using the RADEX program (van der Tak et al.~2007) to calculate LVG models with T$_K=12$~K , we found that the \nthp$J=1\rightarrow0$ emission is a fairly linear function of \nthp column density for densities of 10$^5$ to 10$^{6.3}$~cm$^{-3}$. Below density 10$^5$~cm$^{-3}$, the \nthp column density needs to increase rapidly with decreasing density to emit a fixed line temperature. This means that the emission efficiency of the gas per unit path length decreases rapidly.
In the context of a filament with fixed diameter, the observable emission will decrease unless the \nthp fractional abundance increases rapidly.

The chemistry of \nthp favors a drop in \nthp abundance with decreasing density 
(Bergin et al.~2001, 2002), which translates into a line emission drop. CO destroys \nthp, so in regions less dense than the CO depletion threshold (2-6$\times10^4$~cm$^{-3}$, e.g., Tafalla et al.~2002), the abundance of \nthp decreases. For estimated \nthp abundances of 3-8$\times10^{-10}$ (e.g., Shirley et al.~2005), and a filament path length of 0.03~pc, gas at a density of 10$^5$ cm$^{-3}$ produces \nthp$J=1\rightarrow0$ lines of 1.5-4 K (at the observed FWHM of 0.35\kms), comparable to the strongest observed lines. Hence the observations are consistent with the \nthp emission accentuated by the increase in density and abundance. 
The \nthp emission is therefore found as filamentary structures that are narrower than the dust filaments. Two alternatives are suggested. First, it could be that Herschel resolution is not enough to resolve the substructure inside filaments. Second, it could be that real filaments are broader than observed in \nthp, but density, temperature and abundance effects contribute to the different width measurements.

The above discussion relates to dust filaments composed of single \nthp filaments; there are other broad dust filaments composed of several narrower \nthp filaments. For example, the FSE filament is composed of three \nthp filaments in the southern part that are positioned quasi-parallel to each other and extend over the entire width of the dusty filament (see Figure \ref{moms} and Figure \ref{intcuts}b). A similar morphology is seen in the FE filament, which runs parallel to two shorter \nthp filaments south of it. This is analogous to the bundles found by Hacar et al. (2013) within filaments in a region of the Taurus molecular cloud, which were only resolved in radial velocity (they lacked the resolution to spatially resolve them). The structures presented here are spatially resolved and their paths clearly distinguished and, unlike the bundles found by Hacar et al., most of the quasi-parallel \nthp filaments within a dust filament have similar radial velocities.
 
These results indicate that the dust filaments observed by {\it Herschel} do not always represent one uniform structure within physical filaments, but perhaps are related filaments along the line of sight that project into the appearance of a single filament.
A more robust intensity profile analysis along with further high angular resolution line and continuum observations of more dust filaments in other regions are needed in order to determine if filaments have a common width and the frequency of subfilaments within filaments. 

\subsection{Velocity Gradients in Filaments}
The CLASSy \nthp data also probe the gas kinematics, therefore adding a dynamical perspective to the analysis of these structures. An analysis of the line-of-sight velocity centroid map reveals that the filaments present smooth line-of-sight velocity variations. Some filaments display slight velocity gradients along their length (e.g., FSE and FE; Figure \ref{serps}) while others do not (e.g., FNW). The strength of the observed gradients ($\sim1$~\kms~pc$^{-1}$), corresponds to a crossing time of about a million of years.
Line-of-sight velocity gradients along molecular cloud filaments have been identified in Serpens South and in other regions as well (e.g., Peretto et al.~2014). 
For example, Kirk et al. (2013) also detected a velocity gradient in FSE (0.9\kms~pc$^{-1}$ assuming $d=415$~pc), which is consistent with our measured FSE gradient of 0.8\kms~pc$^{-1}$.
In many cases, these gradients have been interpreted as evidence of material flowing inside the filaments, infalling toward the hubs or filamentary junctions -- regions of active star formation. The typical velocity gradient ranges between 0.2-2~\kms~pc$^{-1}$, coinciding with our estimate of the gradients along the FSE and FE filaments in Serpens South. Although the observed velocities are consistent with those expected from gravitational infall, given the current data this is not the only possible (nor necessarily the preferred) explanation for the observed gradients. 
Most importantly, not all filaments in Serpens South show gradients along their length (see for instance the FNW filament in Figure \ref{serps}), contrary to what is expected if the gravitational pull from the relatively massive hub is the main cause for the gradient. Moreover, even filaments with gradients do not show the symmetrical high velocity infall signature (e.g., Filament A of Figure 2 in Moeckel \& Burcket~2014). A simple alternative for a gradient in radial velocity along a filament's length is that it is the mere projection effect in an elongated structure, perhaps originating from cloud-scale gas turbulence (N. Moeckel private communication). 
In this scenario, parsec-scale turbulent eddies of gas can produce smooth radial velocity gradients in structures with sizes comparable to the eddies. Another possibility is that the velocity variations in the filaments are due to the convergence of two or more flows (Csengeri et al.~2011), and even more sophisticated alternative can include oscillations inside the filament of the material produced by over-densities or starless cores within them (Hacar et al.~2013).

\begin{figure}
\begin{center}
\includegraphics[scale=0.85]{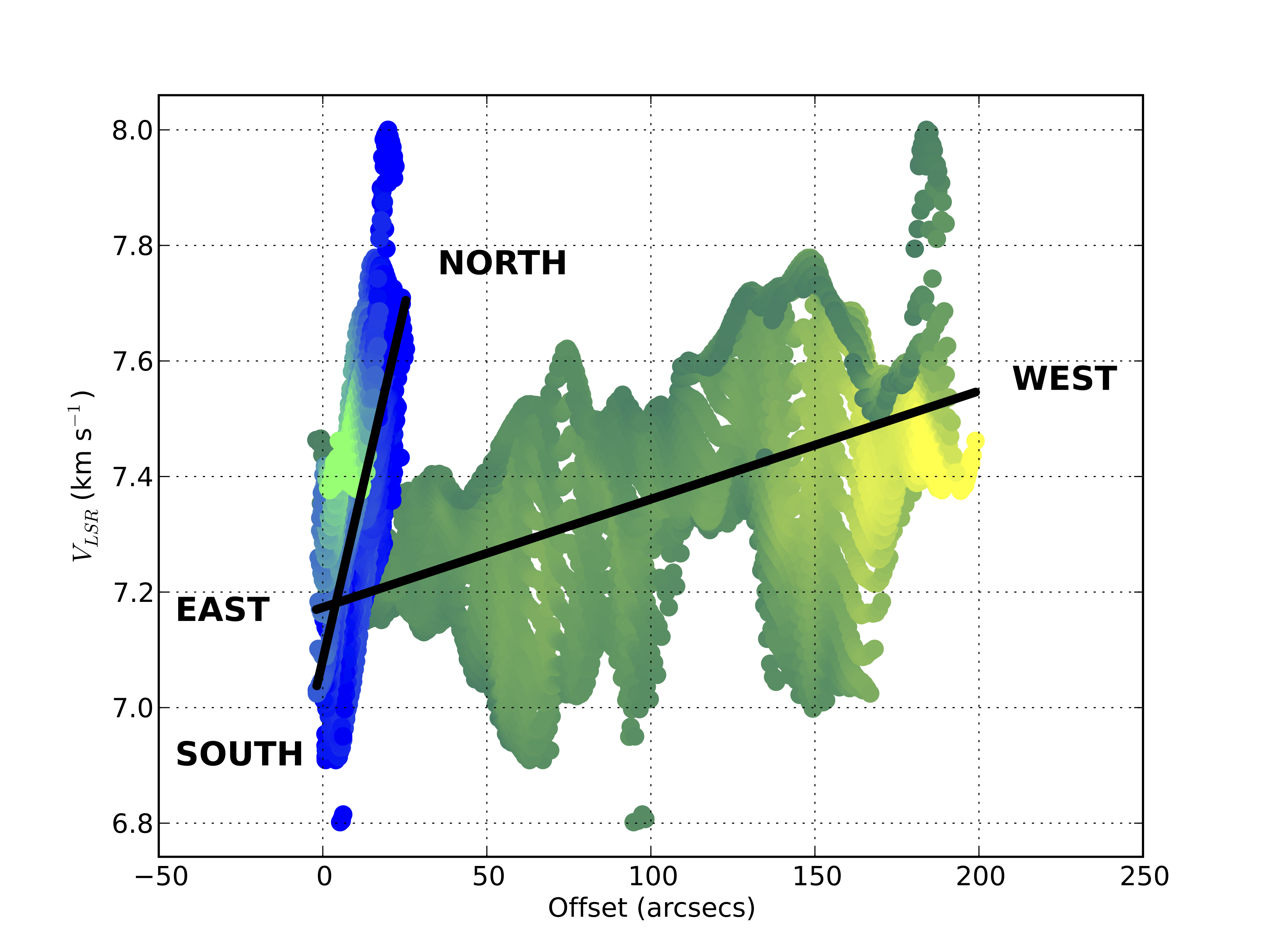}
\caption{Position-velocity plot showing the velocity gradients along (green-to-yellow intensity scale) and across (blue-to-green intensity scale) the FE filament. Each dot corresponds to one pixel of a region encasing this filament (see Figure \ref{moms}b). The straight lines are linear fits to each group of points.
}
\label{gradfe}
\end{center}
\end{figure}

While some velocity gradients are parallel to the main length of the filament, the CLASSy \nthp data also show the existence of velocity gradients perpendicular to several filaments. These gradients are, in several cases, detected along the entirety of the filament. The velocity centroid map of the FNW and FE filaments display these gradients (Figure \ref{serps} and Figure \ref{moms}b). Figure \ref{gradfe} shows both the gradient perpendicular to the FE filament (11.9~\kms~pc$^{-1}$) and the gradient along the filament (0.9~\kms~pc$^{-1}$). Remarkably, one gradient is an order of magnitude larger than the other, indicating that the gas crossing time across the filament is ten times shorter than along the filament (note that gas crossing time is estimated as the time that takes a given gas parcel to travel along or across the filament); this suggests the more dynamical nature of the velocity gradient across (perpendicular to) the filament than the one along (parallel to) the filament. 

To better understand perpendicular line-of-sight velocity gradients in a handful of filaments in Serpens South and other clouds as well, the data are being analyzed and compared against numerical simulations of filament formation via self-gravity in gas that has already been compressed into a locally planar structure by supersonic converging flows (e.g. Gong \& Ostriker 2011, Che-Yu \& Ostriker~2014); the results will be published elsewhere (Mundy et al. in prep.). In these simulations, the velocity gradients are the result of the in-plane motions associated with forming and growing the filament. Gradients across filaments have already been observed in the DR21 filament (Schneider et al.~2010), but in that case, the filament width is about 1~pc, which is more than a factor of 10 wider than the filaments we are analyzing in Serpens South. Therefore, the nature of the gradients across the Serpens South FE and FNW filaments is likely different from those found in DR21.

More generally, gradients across filaments can be thought of as projection effects of kinematics (due to turbulence, self-gravity or other forces; Schneider et al. 2010; Che-Yu \& Ostriker 2014) within a flat structure inclined with respect to the plane of the sky (but see Li \& Goldsmith 2012). An alternative explanation is that two or more narrower filaments with slightly different radial velocities partially overlap on the plane of the sky forming the perpendicular velocity gradient. In any case, these perpendicular gradients are not consistent with the picture in which gradients in filaments trace the flow of material along filaments toward the hub, as the local dynamical evolution will occur on a much faster time than the inward flow timescale.


\section{SUMMARY AND CONCLUSIONS}
We have reported new CARMA \nthp($J=1\rightarrow0$) observations from the CLASSy project toward the Serpens South molecular cloud, covering a region of 3.6~pc$^2$ at a spatial resolution of 7$\arcsec$ and a velocity resolution of 0.16~\kms. The data impose new observational constraints to the origin and evolution of molecular cloud filaments:

\begin{itemize}
\item \nthp filament widths are two to three times narrower than the widths obtained from {\it Herschel} data for the same filaments. We show that this could arise from the density and chemical selectivity of the \nthp emission but it is also possible that the Herschel images are not resolving substructure in the filaments.
\item In a number of cases in Serpens South, single {\it Herschel} filaments are composed of two to three narrower \nthp filaments, often positioned quasi-parallel. This is consistent with these filaments being comprised of separate structures in projection and suggests that filaments in star-forming regions are likely more complex than shown by Herschel observations.
\item Line-of-sight velocity gradients along the filament length are not seen in every filament. When observed they are of order $\sim1$\kms~pc$^{-1}$, consistent with previously detected gradients along filaments in other molecular clouds. The interpretation of these gradients as flows of gas towards the central cluster of protostars is not unique; such gradients can also occur in scenarios where filaments are created by large scale turbulence.
\item We have observed line-of-sight velocity gradients perpendicular to the FE and FNW filaments. This is the first time a gradient across a filament is detected in scales of $\sim 0.03$~pc. The gradients across are steeper than the gradients along these filaments and this adds a new constraint to theoretical models explaining the formation of molecular cloud filaments. These new results are more consistent with a picture where local collapse of filaments is taking place, rather than the scenario where gas flows along filaments feeding material into the central hub.

\end{itemize}

\acknowledgments
We thank all members of the CARMA staff that made these observations possible. 
CLASSy was supported by NSF grants  AST-1139990 (University of Maryland) and AST-1139950 (University of Illinois). Support for CARMA construction was derived from the Gordon and Betty Moore Foundation, the Kenneth T. and Eileen L. Norris Foundation, the James S. McDonnell Foundation, the Associates of the California Institute of Technology, the University of Chicago, the states of Illinois, California, and Maryland, and the National Science Foundation. Ongoing CARMA development and operations are supported by the National Science Foundation under a cooperative agreement, and by the CARMA partner universities. 



\end{document}